\documentclass{PoS}

\title{Status of nucleon structure calculations with 2+1 flavors of domain wall fermions}

\ShortTitle{Status of nucleon structure calculations with 2+1 flavors of domain wall fermions}

\author{\speaker{Meifeng Lin}%
         \thanks{Current address:  Center for Computational Science, Boston University, 3 Cummington Street, Boston, MA 02215, USA.}\\
         for the RBC and UKQCD Collaborations \\
        Department of Physics, Sloane Laboratory, Yale University, New Haven, CT 06520, USA \\
        RIKEN Brookhaven Research Center, Brookhaven National Laboratory, Upton, NY 11973, USA\\

        E-mail: \email{meifeng@caa.columbia.edu}}

\newcommand{\mres}{m_{\rm res}}
\newcommand{\be}{\begin{equation}}
  \newcommand{\ee}{\end{equation}}
\newcommand{\bea}{\begin{eqnarray}}
  \newcommand{\eea}{\end{eqnarray}}

\newcommand{\Dslash}{\rlap{/}\kern-2.0pt D}

\abstract{We report the status of our nucleon structure calculations with 2+1 flavors of domain wall fermions on the RBC-UKQCD $32^3\times$64 gauge ensembles with the Iwasaki+DSDR action. These ensembles have a fixed lattice scale of 1/a = 1.37 GeV, and two pion masses of about 170 and 250 MeV. Preliminary results for the isovector electromagnectic form factors and their corresponding root-mean-squared (r.m.s.) radii will be presented.}

\FullConference{The 30th International Symposium on Lattice Field Theory\\
                 June 24 - 29,  2012\\
                 Cairns, Australia}

\begin{document}

\section{Introduction}
Understanding the internal structure of nucleons has long been a central mission of the theoretical and experimental nuclear physics research. Not only nucleons are the building blocks of the visible matter, but also are they the essential probes into dynamics of the strong interaction, giving us glimpse of how strong interaction works to bind the quarks and gluons together to form protons, neutrons and other hadronic matter. Of particular interests are the nucleon form factors and structure functions, which allow us to study the size, shape and spin structure of the nucleons. Due to the non-perturbative nature of the strong interaction, in which the coupling constant becomes too strong at low energies for perturbatiion theory to be reliable, first-principle calculations of these quantities are best done by lattice QCD approaches. While in the past decade or so the numerical techniques of such calculations have been improved greatly, some results are still plagued with systematic errors that remain to be understood, among which chiral extrapolation, finite volume and excited-state contamination effects have been the most actively investigated ones recently.  Most notably, lattice results for the nucleon axial charge, $g_A$, are consistently lower than the experimental value by 10\%-20\%. Similar behavior has also been observed in the results for the isovector Dirac charge radius. While some studies indicated finite volume effects to be the culprit~\cite{Yamazaki:2008py, Hall:2012qn}, others found that excited-state contaminations may contribute significantly to the discrepancy~\cite{Capitani:2012gj, Green:2012ud}. Of course, almost all the lattice simulations to date have unphysically heavy pion masses, and undetermined errors associated with chiral extrapolations to the physical point may constituent a big part of the discrepancy as well. 

It is thus to our interest to perform lattice calculations in a large volume, with a sufficiently large source-sink separation to suppress excited-state contaminations and at pion masses as close to the physical point as possible. With this in mind, we did our calculations on the 2+1 flavor domain wall fermion gauge configurations \cite{Arthur:2012yc} generated by the RBC and UKQCD collaborations with a spatial volume of about $(4.7 \mathrm{fm})^3$, and two pion masses of about 170 and 250 MeV. As we will discuss later, we also set the source-sink separation to be about 1.3 fm, which is among the largest used in nucleon three-point functions in recent lattice calculations. In these proceedings, we will focus on the calculation of the nucleon electromagnetic form factors, represented by the Dirac and Pauli form factors $F_1^N(Q^2)$ and $F_2^N(Q^2)$. Calculations of the nucleon axial charge and strangeness content on the same gauge ensembles are discussed in separate presentations~\cite{Ohta:2012, Jung:2012} at this conference. 

The Dirac and Pauli form factors are defined through the matrix element
\be
\langle N(p') | J^N_\mu(x) | N(p) \rangle = e^{i (p' - p)\cdot x} \overline{u}(p') \left [ \gamma_\mu F_1^N(Q^2) + i \sigma_{\mu\nu} \frac{q_\nu}{2M_N} F_2^N(Q^2) \right ] u(p), 
\ee
where $p$ and $p'$ are the momenta of the incoming and outgoing nucleons, respectively, and $Q^2 = -q^2 = -(p'-p)^2$. 
It is convenient to define the isovector current $V_\mu(x) \equiv J_\mu^p(x) - J_\mu^n(x) = \overline{u}(x)\gamma_\mu u(x) - \overline{d}(x) \gamma_\mu d(x)$, with the corresponding isovector Dirac and Pauli form factors, $F_1^{p-n}(Q^2)$ and $F_2^{p-n}(Q^2)$.

\section{Details of Calculations}
The calculations were performed on the $32^3\times64$ gauge configurations with 2+1 flavors of domain wall fermions (DWF) and the Iwasaki gauge action with the dislocation-suppressing-determinant ratio (DSDR) at $\beta = 1.75$.  The number of sites in the fifth dimension for DWF was set to be $L_s=32$, resulting in a residual mass of $am_{\rm res} = 0.001842(7)$. Ensembles with two different light dynamical quark masses were used, with $am_l=0.001$ and $0.0042$, corresponding to a pion mass of roughly 170 MeV and 250 MeV, respectively. The strange quark mass was fixed to be $am_s = 0.045$, which is close to the physical strange quark mass.  The lattice cutoff was determined to be $a^{-1}=1.37(1)$ GeV, or $a \approx 0.146$ fm. These gauge configurations were generated jointly by the RBC and UKQCD collaborations, and their basic ensemble properties can be found in Ref.\cite{Arthur:2012yc}. 

We used the standard proton interpolating operator
\be
\chi_S(x) = \epsilon_{abc} \left ( [u_a^S(x)]^TC\gamma_5d_b^S(x) \right ) u_c^S(x),
\ee
where the script $S$ denotes the quark field smearing. In our calculation, we used Gaussian-smeared operator for the sources, and both Gaussian-smeared and local operators for the sinks of the two-point correlation functions, which we write as
\be
C_S(t-t_{src}, p) = \sum_{\vec{x}} e^{i\vec{p}\cdot\vec{x} }\mathrm{Tr}\left [{\cal P}_4\langle 0|\chi_S(\vec{x},t)\overline{\chi}_G(\vec{0},t_{src})|0\rangle \right ].
\ee
The three-point functions always have Gaussian smearing at both the source and the sink. The current operator, of course, remains local. We define our three-point functions as the following
\be
 C_{J_\mu}^{\mathcal{P}_\alpha} (t)= \sum_{\vec{x},\vec{z}} e^{i\vec{q}\cdot\vec{z} }\mathrm{Tr}[\mathcal{P}_\alpha\langle 0 | \chi_G(\vec{x},t_{snk})J_\mu(\vec{z},t)\overline{\chi}_G(\vec{0},t_{src})|0\rangle ],
\ee
where $J_\mu(x) = \overline{q}(x)\gamma_\mu q(x)$, $\mathcal{P}_4 = (1+\gamma_4)/2$ and $\mathcal{P}_{53}=(1+\gamma_4)\gamma_5\gamma_3/2$. 

Since the nucleon correlation functions decay quickly in the time direction, we chose to use four different source locations on a gauge configuration to take advantage of the large time extent. For both ensembles, we placed the sources at $t_{src} = 0, 16, 32$ and $48$, while the spatial coordinates of the sources were fixed at the origin, that is, $(x,y,z)_{src} = (0, 0, 0)$. The calculations were done on gauge configurations separated by every eight molecular dynamics units. On the $am_l=0.001$ ensemble, 103 gauge configurations were used, while on the $am_l=0.0042$ ensemble, 165 gauge configurations were analyzed, making the total number of correlation functions 412 and 660 respectively. These parameters are summarized in Table~\ref{tab:params}. 

 \begin{table}[ht]
        \centering
        \begin{tabular}{ccccccccc}
    
          \hline
          \hline
          $am_l$ & $am_s$ & $L^3\times T$ & $L_s$ & {$m_\pi$ [MeV]} & {$m_\pi L$} & $a\mres$ & \# of configs. & \# of meas.\\
          \hline
          0.001 & 0.045 & $32^3\times64$ & 32 & {170} & {4.0} & 0.0018 & 103 & 412 \\
          \hline
          0.0042 & 0.045 & $32^3\times 64$ & 32 & {250} &{5.8} & 0.0018 & 165 & 660 \\
           \hline
          \hline
        \end{tabular}
        \caption{Details of the calculations.}
        \label{tab:params}
      \end{table} 

For the three-point functions to have sufficient overlap with the nucleon ground state, we need to pay particular attention to both the nucleon sources and the source-sink separations. The optimal choice of the source-sink separation depends on the choice of the nucleon source. As mentioned above, we used the Gaussian smearing for our sources. As pointed out in Ref.\cite{Syritsyn:2009mx}, gauge link smearing can help suppress the gauge noice. Hence we also applied APE smearing to the gauge links in the construction of the nucleon sources. Thus our proton operator consists of smeared quark fields defined as 
\be
\psi_{G,n_G}(x) = \left (\psi(x)-\frac{\sigma^2}{4n_G} \sum_{i=1,2,3} \left [ 2\psi(x) - U_i^\prime(x)\psi(x+\hat{i}) - U_i^{\prime \dagger}(x-\hat{i})\psi(x-\hat{i})\right ] \right )^{n_G},
\ee
where $U_i^\prime(x)$ is the APE smeared gauge link, obtained by applying $N = 25$ steps of the following transformation, 
\bea
U_\mu^\prime(x) &=& (1-c) U_\mu(x) + \frac{c}{6} \sum_{\nu = 1, \nu\neq \mu}^4 \left [ U_\nu(x)U_\mu(x+\hat\nu)U_\nu^\dagger(x+\hat\mu) \right. \\
&+& \left . U_\nu^\dagger(x-\hat\nu) U_\mu(x-\hat\nu)U_\nu(x-\hat\nu+\hat\mu) \right ].
\eea
We set $\sigma=6.0$, $n_G=70$ and $c=0.4$, which we found give a nucleon operator that has a good overlap with the ground state~\cite{Ohta:2010sr}. 

The source-sink separation, $t_{snk} - t_{src}$, was set to be 9 lattice units, amounting to about 1.3 fm in physical units. As we will see in Section~\ref{sec:results}, this choice gives us nice plateaus to determine the form factors for the nucleon ground state. We fixed the sink momentum to be $\vec{p}'=(0,0,0)$, while the momenta for the operator, $\vec{q} \equiv\frac{2\pi}{L}  \vec{n} $, takes all the possible values with $|\vec{n}|^2 \leq 4$. 

\section{Preliminary Results}\label{sec:results}
We follow the steps as described in \cite{Yamazaki:2009zq} to obtain form factors from the nucleon three-point and two-point functions. We first average over the four source locations before performing the standard jackknife analysis. No further blocking has been done yet, thus it is possible that the errors are underestimated. As we have only included connected contractions in our calculations, we will only present results for isovector quantities, in which the disconnected-diagram contributions are canceled. 

\begin{figure}[htbp]
\centering
\includegraphics[width=0.42\textwidth,clip]{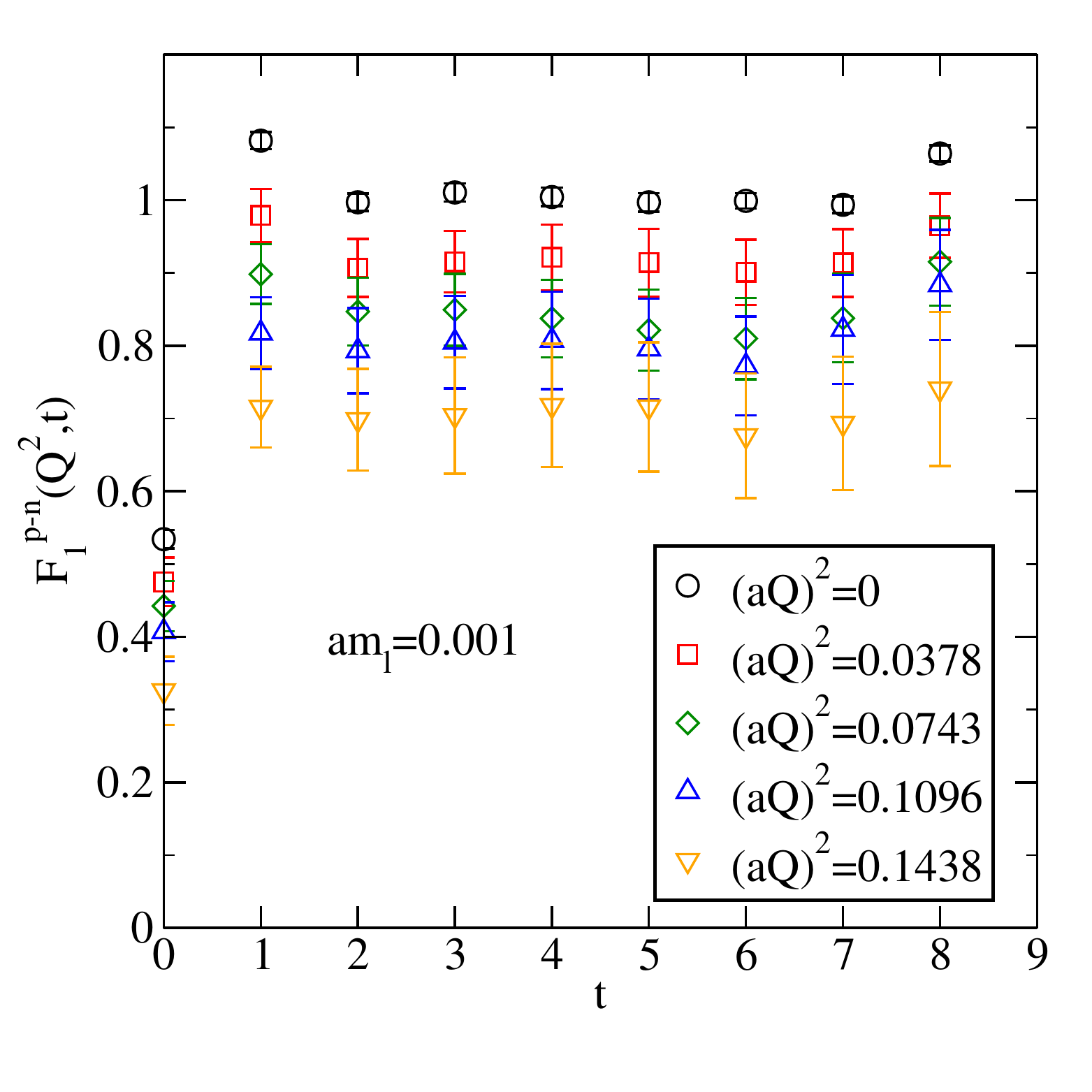} 
\includegraphics[width=0.42\textwidth,clip]{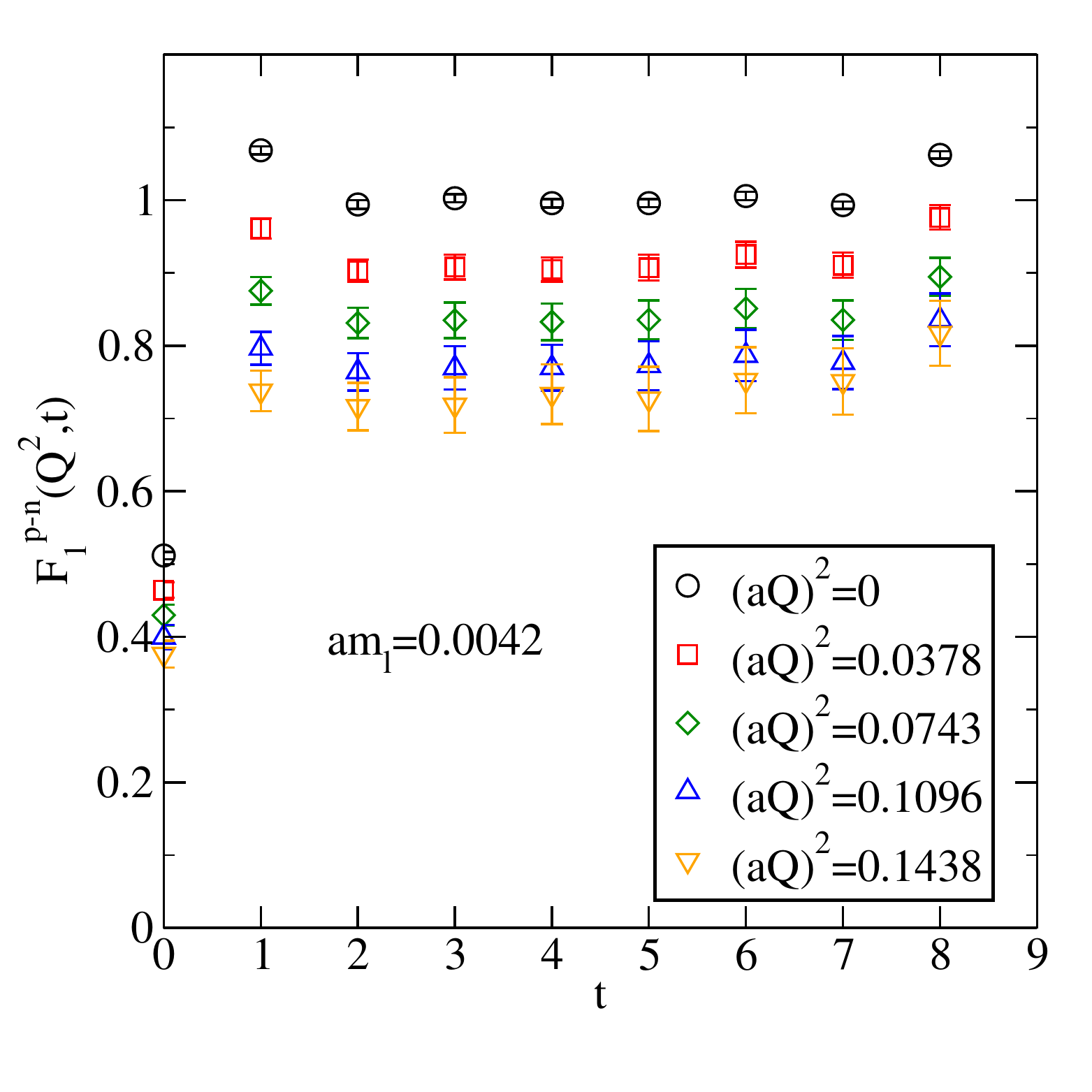}
\caption{The plateaus for the isovector Dirac form factor, $F_1^{p-n}(Q^2)$, from the two ensembles with $am_l=0.001$ (left) and $am_l=0.0042$ (right). \label{fig:F1_t} }
\end{figure}

\begin{figure}[htbp]
\centering
\includegraphics[width=0.42\textwidth,clip]{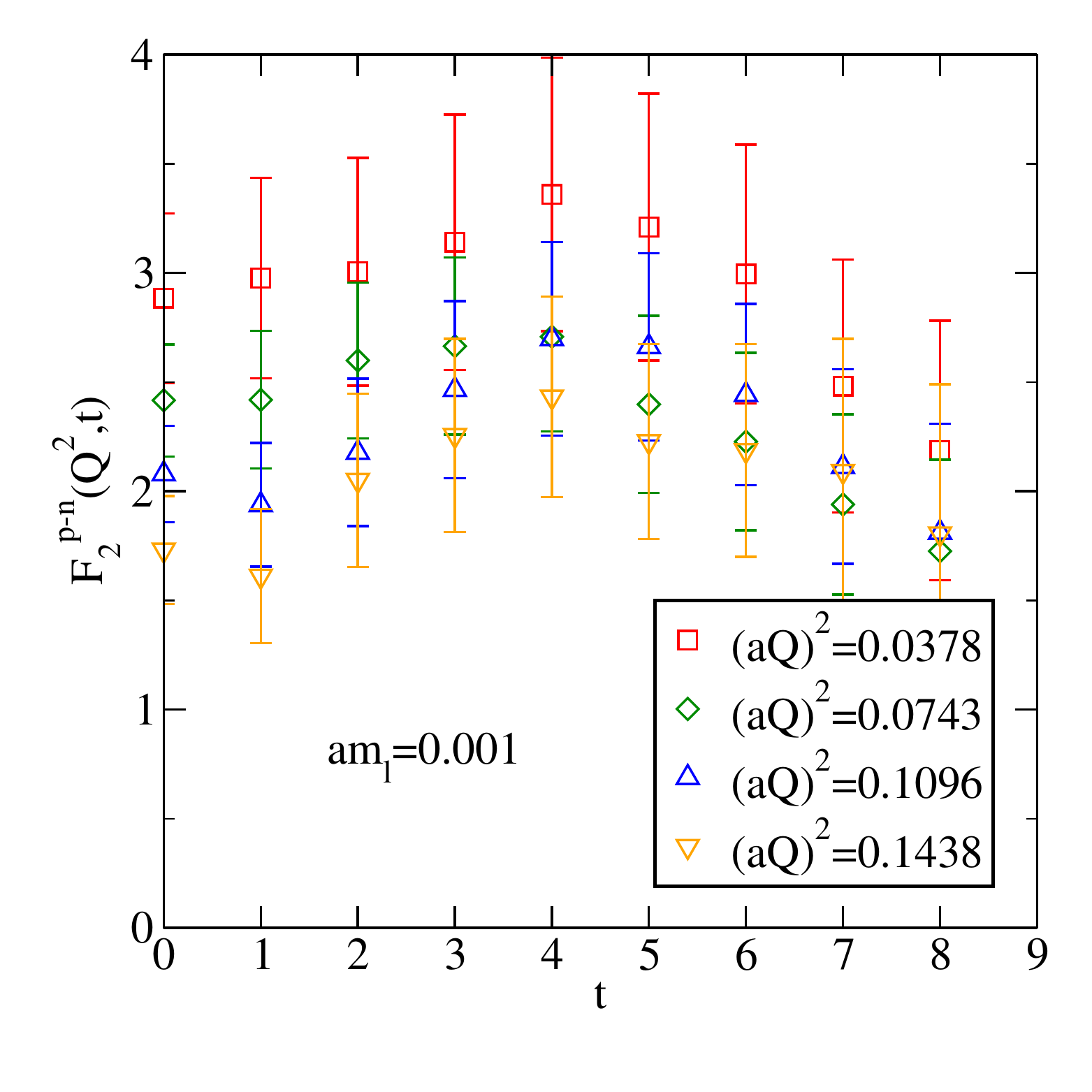} 
\includegraphics[width=0.42\textwidth,clip]{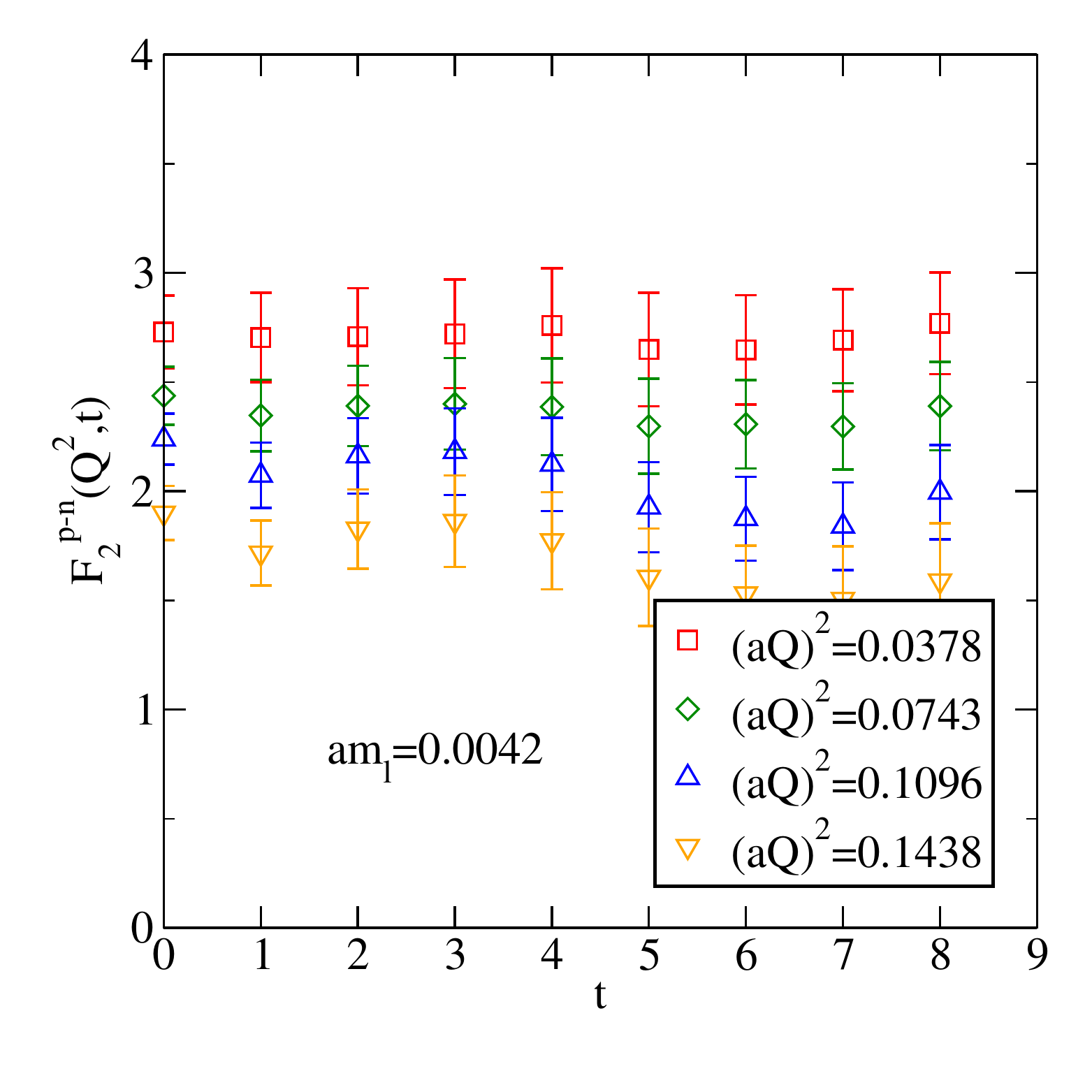} 
\caption{The plateaus for the isovector Pauli form factor, $F_2^{p-n}(Q^2)$, from the two ensembles with $am_l=0.001$ (left) and $am_l=0.0042$ (right). \label{fig:F2_t}
}
\end{figure}

\begin{figure}[htbp]
\centering
\includegraphics[width=0.42\textwidth,clip]{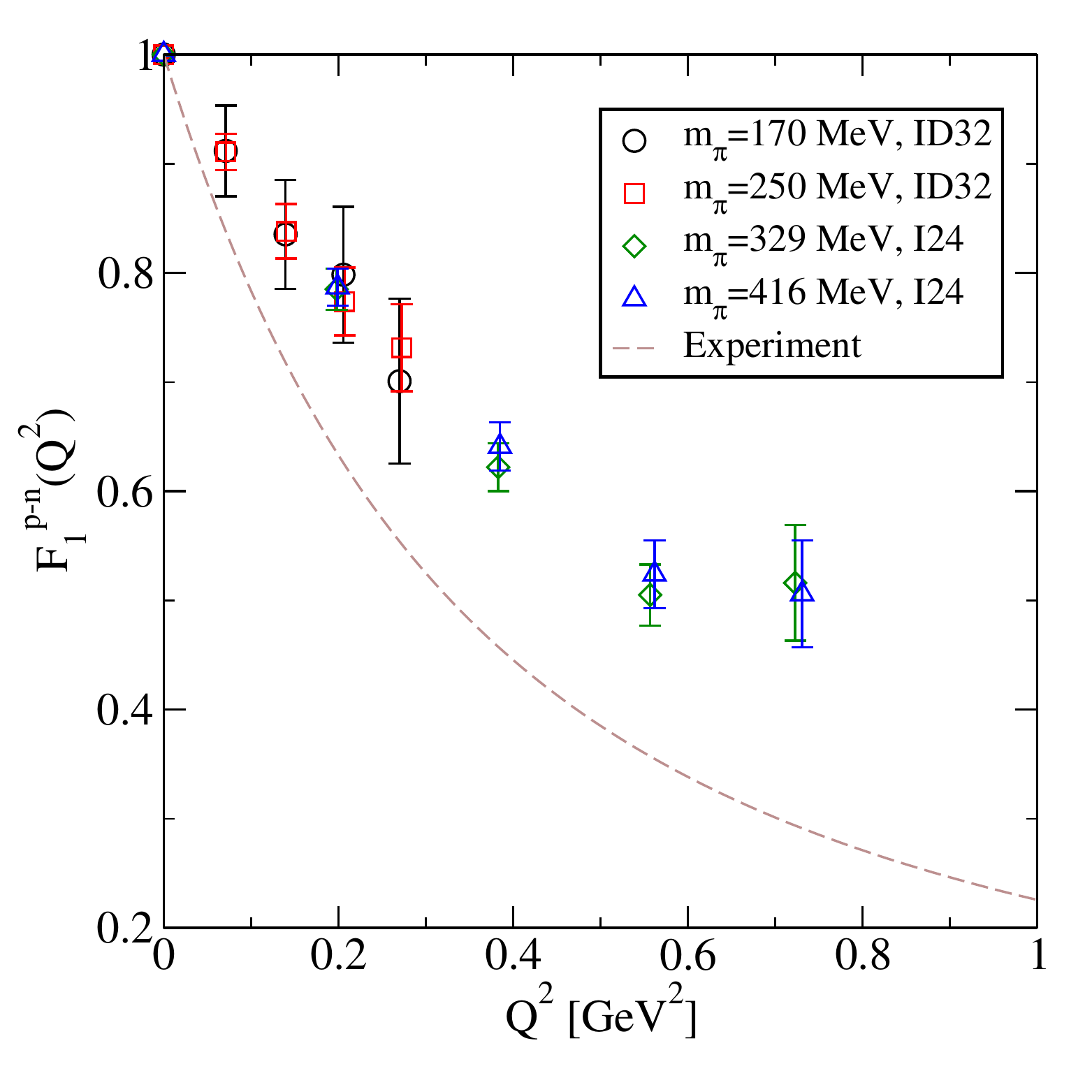} 
\includegraphics[width=0.42\textwidth,clip]{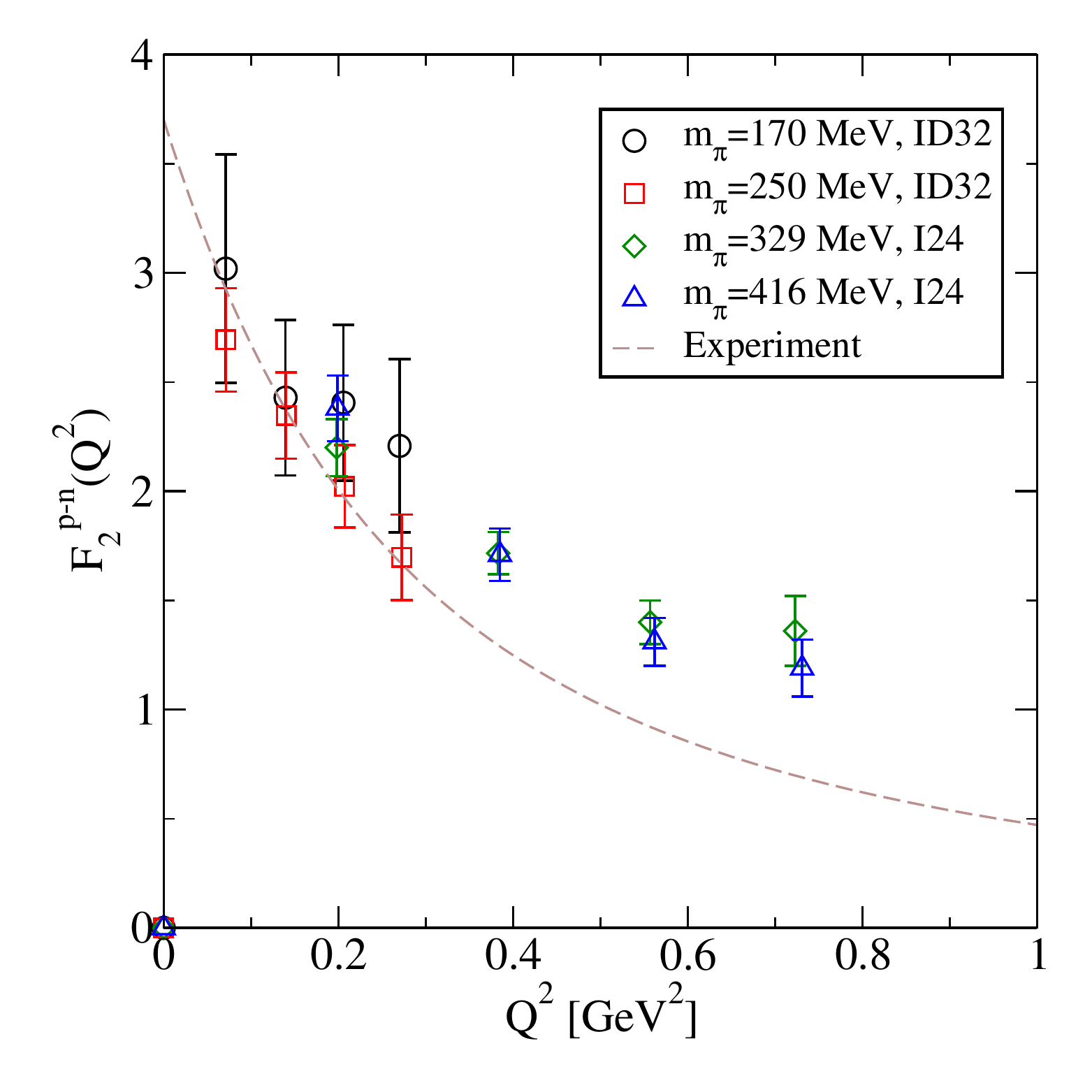} 
\caption{$Q^2$ dependence of the isovector Dirac (left) and Pauli (right) form factors, $F_1^{p-n}(Q^2)$ and $F_2^{p-n}(Q^2)$. The results from the calculation discussed in these proceedings are labeled as ``ID32''. Results labeled as ``I24'' are from Ref.~\cite{Yamazaki:2009zq}. \label{fig:F1F2_Q}}
\end{figure}

  
Figs.~\ref{fig:F1_t} and \ref{fig:F2_t} show the quality of the plateaus for the isovector Dirac and Pauli form factors, respectively. For $F_1(Q^2,t)$, the plateaus settle in after the first two time slices for all $Q^2$  values, while for $F_2(Q^2,t)$,  there are some fluctuations in the middle of the time slices for $am_l=0.001$. For simplicity, we choose the fit range to be $t=[2,7]$ for all quantities. With the large statistical errors for the $am_l=0.001$ ensemble, the systematic errors that may result from the choice of fit ranges should be subleading compared to the statistical errors quoted. In the future when more statistics are available, the choice of the fit ranges will be re-examined. The results, labeled as ID32, for $F_1(Q^2)$ and $F_2(Q^2)$ are shown in Fig.~\ref{fig:F1F2_Q}. Previous results~\cite{Yamazaki:2009zq} from the calculation on the $24^3\times64$ lattices with a lattice cutoff of $a^{-1}\approx1.73$ GeV are also shown for comparison  (I24). The pion mass dependence for $F_1(Q^2)$ is very mild for all the ensembles, and the $Q^2$ dependence is not as steep as the Kelly parametrization of the experimental data~\cite{Kelly:2004hm}, shown as the dashed lines in Fig.~\ref{fig:F1F2_Q}. The ID32 results for $F_2(Q^2)$, however, seem to agree with the experiment at smaller $Q^2$ values, which translates into promising trend for the Pauli radii that we will discuss next.

\begin{figure}[htbp]
\centering
\includegraphics[width=0.42\textwidth,clip]{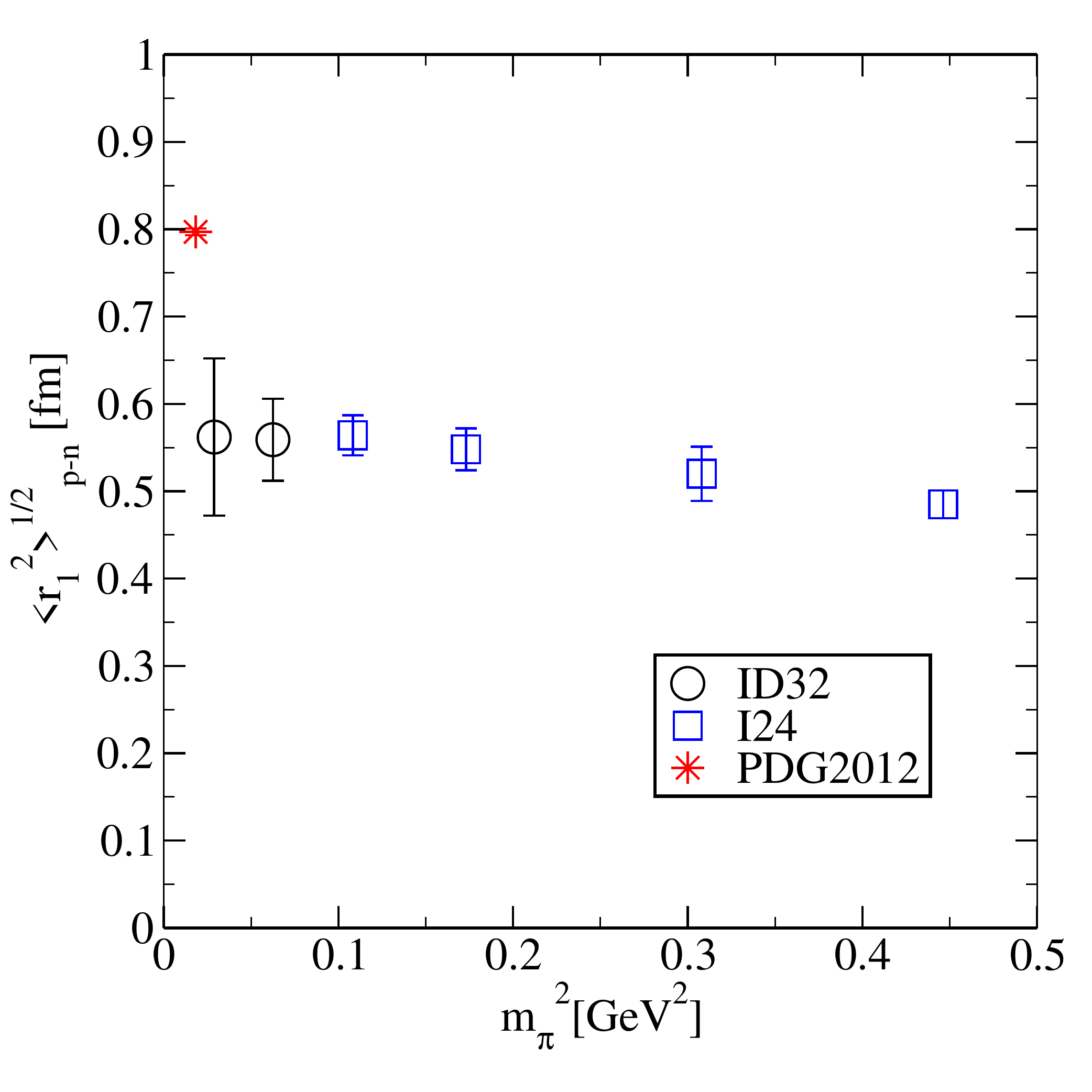}
\includegraphics[width=0.42\textwidth,clip]{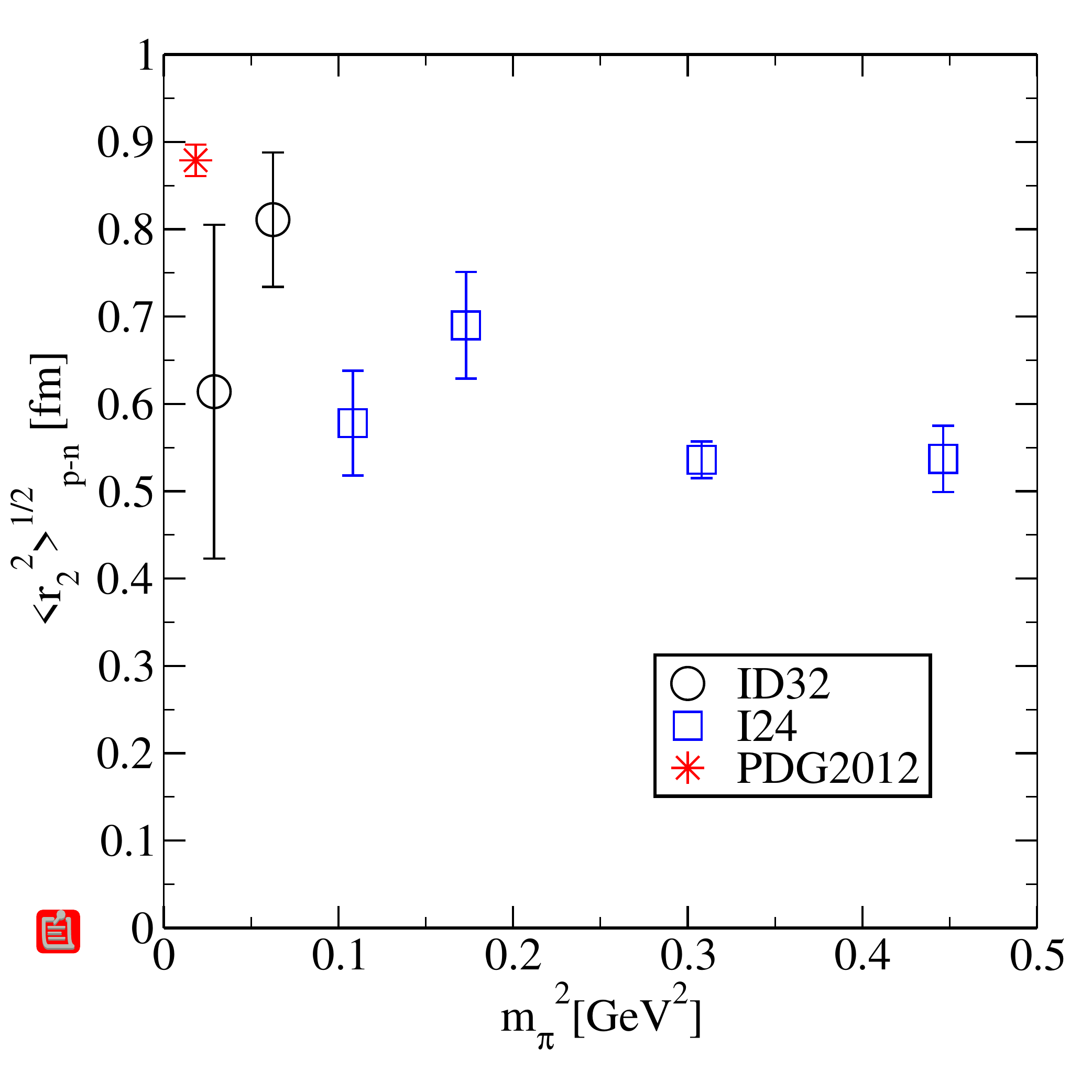}
\caption{Pion mass dependence of the isovector Dirac (left) and Pauli (right) r.m.s radii, $\langle r_1^2 \rangle^{1/2}_{p-n}$ and $\langle r_2^2 \rangle^{1/2}_{p-n}$. The results from the calculation discussed in these proceedings are labeled as ``ID32''. Results labeled as ``I24'' are from Ref.~\cite{Yamazaki:2009zq}. \label{fig:radii}}

\end{figure}

We also determine the root-mean-squared (r.m.s) radii from the $Q^2$ dependence of the form factors. The mean-squared radii are defined as 
\be
\langle r_i^2 \rangle = -\frac{6}{F_i(0)} \frac{\partial F_i(Q^2)}{\partial Q^2}|_{Q^2 = 0}, \,\, i = 1, 2. 
\ee
We fit $F_1(Q^2)$ and $F_2(Q^2)$ to the empirical dipole form of $F_i(Q^2) = \frac{F_i(0)}{\left ( 1+ {Q^2}/{M_i^2} \right)^2}$, from which we get $\langle r_i^2 \rangle = \frac{12}{M_i^2}$. For $F_1(Q^2)$, as the form factors are normalized to 1 at $Q^2=0$, we only allow $M_1$ to be the free parameter in the fit, while for $F_2(Q^2)$, both $F_2(0)$ and $M_2$ are free parameters. The results for the r.m.s. radii, along with results from an earlier, I24, calculation, are shown in Fig.~\ref{fig:radii}. For $\langle r_1^2\rangle^{1/2}$ (Fig.~\ref{fig:radii}, left), all the results seem to fall on a straight line with respect to $m_\pi^2$, and can be linearly extrapolated to a value which is 20\%-30\% lower than the experiment\footnote{A measurement using muonic hydrogen found a much smaller charge radius~\cite{Pohl:2010zza}, which was not included in the PDG average.}~\cite{Beringer:1900zz}. This linear behavior has been observed in other lattice calculations too, as summarized in \cite{Yamazaki:2009zq}. Recently, the authors in ~\cite{Green:2012ud} found that the lack of curvature of the results for $\langle r_1^2\rangle^{1/2}$ may be due to excited-state contaminations. Whether this is the case for our calculation still awaits further investigations. 

The results for $\langle r_2^2\rangle^{1/2}$ (Fig.~\ref{fig:radii}, right) are more interesting. The lightest point from the ID32 calculation and that  from the I24 calculation both dip down to be away from the experiment. If these two points are not considered, then the results seem to approach the experiment on a smooth curve. One possible explanation is that the two lightest points, with $m_\pi L = 4$ and $4.5$, respectively, may suffer from large finite volume effects. Of course, there are other systematics that may cause this behavior, and further investigations are needed to draw a definitive conclusion. 

\section{Summary and Outlook}
We have reported a calculation of the nucleon isovector electromagnetic form factors on the 2+1 flavor domain wall fermion gauge configurations with a lattice volume of $32^3\times64$ and a lattice cutoff of $a^{-1} \approx 1.37$ GeV. Two pion masses, 170 MeV and 250 MeV, have been studied. We have presented some preliminary results for the isovector Dirac and Pauli form factors, as well as the corresponding r.m.s radii. We continue to see the linear trend in the results for the r.m.s Dirac radius, which, if extrapolated linearly, are about 20\%-30\% below the experiment.  The results for the r.m.s Pauli radius show promising signs of approaching the experiment as the pion mass is decreased, with the exception of two points with the smallest $m_\pi L$ ( from 4 to 4.5), which may be an indication of finite volume effects. To understand these issues conclusively, we will need to increase our statistics, and perform thorough studies of excited-state contaminations, finite volume effects and other systematic errors. We are currently doubling the statistics for both ensembles, and are also applying error-reduction techniques~\cite{Blum:2012uh} to further increase the statistical accuracy of our results. 

\section*{Acknowledgments}
Part of this work used the Extreme Science and Engineering Discovery
Environment (XSEDE), which is supported by National Science Foundation
grant number OCI-1053575. We also thank the RIKEN Integrated Cluster
of Clusters (RICC) for the computer resources used in the
calculations. M.L. is partly supported by National Science Foundation under Grant
No. NSF PHY11-00905. The gauge configurations used in these calculations were generated jointly by the RBC and UKQCD Collaborations. The author thanks Yasumichi Aoki, Tom Blum, Chris Dawson, Taku Izubuchi, Chulwoo Jung, Shigemi Ohta, Shoichi Sasaki, Eigo Shintani and Takeshi Yamazaki for many useful discussions and important contributions. 
\bibliographystyle{h-physrev5}
\bibliography{refs}

\end{document}